\documentclass[12pt]{article}
\usepackage{amsmath}
\usepackage{graphicx}
\usepackage{enumerate}
\usepackage{natbib}
\usepackage{multirow}
\usepackage{longtable}
\usepackage{booktabs}
\usepackage{amssymb}
\usepackage{bm}
\usepackage{mathrsfs}
\usepackage{amsthm}
\usepackage{amsfonts}
\usepackage{subfigure}
\usepackage{floatrow}
\usepackage{epsfig,amssymb,latexsym,verbatim}
\usepackage{epstopdf}
\usepackage{graphics}
\usepackage[english]{babel}
\usepackage[figuresright]{rotating}
\usepackage[dvipsnames]{xcolor}
\usepackage{color}
\usepackage{hyperref}

\usepackage{cases}

\usepackage{caption}
\usepackage{filecontents}
\usepackage{algorithmic}
\usepackage{algorithm}

\newtheorem{theorem}{Theorem}[section]
\newtheorem{corollary}{Corollary}[section]

\newtheorem{proposition}{Proposition}[section]

\newtheorem{remark}{Remark}[section]

\usepackage{url} 

\newcommand{\blind}{0}
\usepackage{xr}
\def\P{\mathbb{P}}
\def\E{\mathbb{E}}

\def\o{{\scriptstyle{\mathcal{O}}}}
\def\O{\mathcal{O}}

\allowdisplaybreaks

\begin{document}

\def\spacingset#1{\renewcommand{\baselinestretch}%
{#1}\small\normalsize} \spacingset{1}

\date{}
\if0\blind
{
  \title{\bf Statistical inference for high-dimensional convoluted rank regression
}
  \author{Leheng Cai\footnotemark[1],    Xu Guo\footnotemark[2] \footnotemark[3],  Heng Lian\footnotemark[4], and Liping Zhu\footnotemark[5]\hspace{.2cm}}
  \maketitle
  \renewcommand{\thefootnote}{\fnsymbol{footnote}}
		\footnotetext[1]{ Department of Statistics and Data Science, 			Tsinghua University}
		\footnotetext[2]{School of Statistics, Beijing Normal University, Beijing, China}
		\footnotetext[4]{Department of Mathematics, City University of Hong Kong, Hong Kong, China}
		\footnotetext[5]{Institute of Statistics and Big Data, Renmin University of China, Beijing, China}
		\footnotetext[3]{We would like to thank the Editor, the Associate Editor, and the three anonymous reviewers for their valuable comments and constructive suggestions, which lead to significant improvements in the paper. Corresponding Author: Xu Guo. Email address: xustat12@bnu.edu.cn.}
} 

\if1\blind
{
  \bigskip
  \bigskip
  \bigskip
  \begin{center}
    {\LARGE\bf Statistical inference for high-dimensional convoluted rank regression}
\end{center}
  \medskip
} \fi

\bigskip
\begin{abstract}
   High-dimensional penalized rank regression is a powerful tool for modeling high-dimensional data due to its robustness and estimation efficiency. However, the non-smoothness of the rank loss brings great challenges to the computation. To solve this critical issue, high-dimensional convoluted rank regression has been recently proposed, introducing penalized convoluted rank regression estimators.  However, these developed estimators cannot be directly used to make inference. In this paper, we investigate the statistical inference problem of high-dimensional convoluted rank regression. The use of U-statistic in convoluted rank loss function presents challenges for the analysis. We begin by establishing estimation error bounds of the penalized convoluted rank regression estimators under weaker conditions on the predictors. Building on this, we further introduce a debiased estimator and  provide its Bahadur representation. Subsequently,   a high-dimensional Gaussian approximation for the maximum deviation of the debiased estimator is derived, which allows us to construct  simultaneous confidence intervals.  For implementation, a  novel bootstrap procedure is proposed and its theoretical validity is also established. Finally, simulation and real data analysis are conducted to illustrate the merits of our proposed methods. 	 
\end{abstract}

\noindent%
{\it Keywords:}  Convoluted smoothing, debiased
method, high-dimensional convoluted rank  regression,  simultaneous inference.

\newpage
\spacingset{1.9} 

\section{Introduction}

Due to the development of modern technology, data sets with high-dimensional predictors are frequently collected. The presence of outliers and heavy-tailed noises is an important feature of high-dimensional data. Statistical procedures that neglect these factors may cause misleading results. To deal with outliers and heavy-tailed noises, many robust procedures are developed for high-dimensional data. Indeed, \cite{fan2017estimation} and \cite{sun2020adaptive} made in-depth investigation about high-dimensional Huber regression. \cite{belloni2011} investigated high-dimensional quantile regression with LASSO penalty \citep{tibshirani1996regression}. \cite{wang2012quantile} studied the nonconvex penalized quantile regression model via SCAD penalty \citep{fan2001variable}. \cite{wang2009weighted} studied the penalized rank regression for low-dimensional data, while \cite{wang2020tuning} further investigated the penalized rank regression in the high-dimensional setting. The penalized rank
regression is not only robust, but also enjoys high estimation efficiency.  Compared with the least squares method, rank regression estimator has a substantial gain in estimation efficiency while maintaining a minimal relative efficiency of $86.4\%$ under arbitrary symmetric error distribution with finite Fisher information \citep{hettmansperger2010robust}. However, there is a big obstacle for practical application of  the rank regression. In fact, the loss function in rank regression is non-smooth,   thus the computation is a major challenge. 

To reduce the computation time of rank regression, 
\cite{zhou2023sparse} introduced a novel convoluted rank regression. The convoluted smoothing technique was introduced by \cite{fernandes2021smoothing} for quantile regression. \cite{he2023smoothed} and \cite{tan2022high} made theoretical investigation deeply of the convoluted quantile regression in high-dimension. Different from convoluted quantile regression, convoluted rank regression does not introduce any smoothing bias. On the other hand, same as the convoluted quantile regression, the convoluted rank loss function in convoluted rank regression is smooth and convex, and thus the computation of rank regression is made easier. 

For the high-dimensional convoluted rank regression, \cite{zhou2023sparse} proposed penalized estimators, and derived their estimation error bounds under the fixed design and the boundedness assumption of the predictors. This adoption of relatively strong assumptions on the predictors is due to the fact that the convoluted rank loss function is of U-statistic nature. Without the fixed design and the boundedness assumption, the theoretical analysis could be very challenging. To extend their results, we first  establish estimation error bounds of the penalized convoluted rank regression estimators under weaker conditions on the predictors.

Besides, the penalized convoluted rank regression estimators in \cite{zhou2023sparse} cannot be directly used to make inference, since the penalized convoluted rank regression estimators with LASSO penalty are generally biased \citep{zhang2008sparsity, zhang2014confidence}. Additionally, the strong oracle property of penalized convoluted rank regression estimators with folded concave penalties in \cite{zhou2023sparse} requires the minimal signal condition, which  may be restrictive and may not be easy to verify. 
In this paper, we investigate the statistical inference problem of high-dimensional convoluted rank regression without the minimal signal condition. A debiased estimator is proposed, along with the corresponding Bahadur representation. Our procedures are inspired by the debiasing technique developed by many authors, see for instance, \cite{zhang2014confidence, van2014, javanmard2014confidence, ning2017general}.

{\color{black}
Recently, there are many developments about the inference of the linear regression coefficients under heavy-tailed noises. To be specific, \cite{loh2021scale} and \cite{han2022robust} developed robust inference methods for the regression coefficients based on the Huber loss. \cite{song2023large} developed a robust multiple
testing procedure to test a large number of general linear hypotheses for the regression coefficients in multivariate linear regression. 
In their paper, 
the dimension of the response variables can diverge exponentially fast with the sample size $n$, but $p/n\to 0$ where $p$ is the dimension of covariates $\bm X$. While we focus on one-dimensional response $Y$, but allow the dimension of covariates diverges exponentially fast with the sample size. \cite{yan2023confidence} considered the inference problem of high-dimensional convoluted quantile regression by adopting the debiasing technique. \cite{zhao2014general} further considered the inference problem of high-dimensional composite quantile regression. Alternatively, there are also some invariance-based inference methods.  \cite{wang2021robust} proposed a robust residual randomization-based test for testing $H_0:\bm a^\top \bm\beta=a_0$, where $\bm a\in\mathbb R^p$ and $a_0\in\mathbb R$ are given in advance. By inverting 
the hypothesis test and letting $\bm a=\bm e_j$, one could construct the confidence interval for an individual coefficient $\beta_j$. \cite{guo2023invariance} proposed a residual-based test for testing $H_0:\bm\beta_{\mathcal{G}}=\bm 0$, where $\mathcal{G}\subseteq \{1,2,\ldots,p\}$. They assumed $p-|\mathcal{G}|<n$, and the fourth moment of the error term exists.

Our paper is different from the above developments in the following aspects. First, our method makes no moment condition for the error term. While most of the above papers require at least the second order moment of the error term exist. See for instance \cite{loh2021scale}, \cite{guo2023invariance} and \cite{song2023large}. Second, a main theoretical difference from the existing analysis of debiased estimators is that the objective function in our paper is not a sum of independent variables but a U-statistic, which requires additional efforts to conduct theoretical analysis. Third, our paper investigates the high-dimensional simultaneous inference problem \citep{zhang2017simultaneous, dezeure2017high, ning2017general, ma2021global, wu2023model} by allowing $|\mathcal{G}|$ be diverging with the sample size $n$. While most of the above papers can only make inference for fixed dimensional parametric vector. High-dimensional simultaneous inference problem requires high-dimensional central
limit theorem and high-dimensional bootstrap \citep{chernozhukov2023high}, which could be more challenging than inference for fixed-dimensional parameters. 

We make a deep investigation about simultaneous inference problem for the convoluted rank regression. Theoretically we derive a high-dimensional Gaussian approximation result for the maximum deviation of the debiased estimator, while methodologically we develop a novel bootstrap procedure to make simultaneous inference. For the Gaussian and bootstrap approximations for high-dimensional U-statistics, \cite{chen2017} and \cite{cheng2022gaussian} made important contributions. For other recent improvements in high-dimensional Gaussian approximation for U-statistics, we refer to \cite{song2019}, \cite{chen2019} and \cite{chen2020jackknife}. However the elegant results in \cite{chen2017} and \cite{cheng2022gaussian} cannot be directly used here, since our U-statistics involve unknown quantities. Our developed methods and theories could also be useful for other problems involving U-statistic type loss functions. 

Further, our result about estimation error bound extends the results in \cite{zhou2023sparse} in two aspects. Firstly in \cite{zhou2023sparse}, they assumed that the predictors are fixed design and bounded uniformly, while in our paper we consider random design setting and assume that the predictors are sub-Gaussian, which is milder than the boundedness assumption. Secondly, we derive the estimation error bounds of penalized estimators with many penalty functions including LASSO, SCAD and MCP in a unified approach. For nonconvex penalty functions, our results show that all local minima within a feasible region have desirable error bounds. These new results are important, because the global minimizers for nonconvex objective functions are not always numerically
 obtained or verifiable in practice.

}

To summarize, we make the following contributions.
\begin{itemize}
\item Firstly, we revisit the penalized convoluted rank regression estimators in \cite{zhou2023sparse} and establish estimation error bounds under weaker conditions on the predictors and with many penalty functions in a unified approach. 
\item Secondly, we proposed a debiased convoluted rank regression estimator and further obtain its Bahadur representation.   
\item Thirdly, we derive a high-dimensional Gaussian approximation result for the maximum deviation of the debiased estimator,  and develop a novel bootstrap procedure with theoretical justification to make simultaneous inference.
\end{itemize}

The rest of this article is organized as follows. In Section \ref{sec2}, we introduce debiased convoluted rank regression estimators and develop its Bahadur representation. In Section \ref{sec3}, we investigate simultaneous inference and introduce a novel bootstrap procedure. We also establish the bootstrap procedure's validity. In Section \ref{sec4}, we present some simulation studies. A real data example is presented in Section \ref{sec:realdata}. Conclusions and discussions are given in Section \ref{sec6}. All technical proofs and additional simulation studies are given in the Supplementary  Material.

\par {\textbf{Notations:}} 
For any vector $ \bm{a} = \left(a_1,\ldots,a_s  \right)^\top \in \mathbb{R}^s $, take $ \left\Vert \bm{a}\right\Vert_r = \left( \left| a_1\right|^r+ \ldots+\left| a_s\right|^r \right)^{1/r}$ for $1 \le r < \infty$, $\left\|\bm a\right\|_0=\sum_{j=1}^{s}\mathbf 1\{a_j\neq 0\}$,  and $ \left\Vert \bm{a} \right\Vert_\infty=\max_{1\leq j\leq s}|a_{j}|$. For any $n\times n$ matrix $\mathbf A=(a_{ij})_{i,j=1}^{n}$, denote its $L_r$-norm as $\left\|\mathbf A\right\|_{r}=\sup_{\left\|x\right\|_r=1}\left\|\mathbf Ax\right\|_{r}$ for $1\leq r \le \infty$, and particularly $\left\|\mathbf A\right\|_1=\max_{1\leq j\leq n}\sum_{i=1}^{n}|a_{ij}| $ and $\left\|\mathbf A\right\|_{\infty}=\max_{1\leq i\leq n}\sum_{j=1}^{n}|a_{ij}|$. Further define $\left\|\mathbf A\right\|_{\max}=\max_{1\leq i,j\leq n}|a_{ij}|$ and $\left\|\mathbf A\right\|_{\ell_1}=\sum_{i=1}^{n}\sum_{j=1}^{n}|a_{ij}|$.  
The sub-Gaussian norm of a sub-Gaussian random variable $X$ is
defined as $\left\|X\right\|_{\psi_2}=\inf\{t>0:\E\exp(X^2/t^2)\leq 2\}$, and the sub-Exponential norm of
a sub-Exponential random variable $Y$ is defined as $\left\|Y\right\|_{\psi_1}=\inf\{t>0:\E\exp(|Y|/t{\color{black})}\leq 2\}$. 
For two sequences $\{a_n\}$ and $\{b_n\}$, we write $a_n\asymp b_n$, $a_n\lesssim b_n$ and $a_n\gg b_n$ to mean
that there exist $c$, $C>0$ such that $0<c\leq |a_n/b_n|\leq C<\infty$ for all $n$,  $a_n\leq  Cb_n$ for all $n$, and $b_n=\o(a_n)$
respectively.

\section{Debiased convoluted rank regression estimator}\label{sec2}
 Assume
that the data $\{\bm X_i,Y_i\}_{i=1}^{n}$ are generated from the following linear
regression model,
\begin{align*}
    Y_i=\bm X_i^\top \bm\beta^*+\epsilon_i,
\end{align*}
where $\{Y_i\}_{i=1}^n$ are one-dimensional responses, $\{\epsilon_i\}_{i=1}^{n}$ are random errors, $\{\bm X_i\}_{i=1}^{n}$ are $p$-dimensional random  vectors with mean zero and covariance matrix $\mathbf\Sigma$, and is independent of $\epsilon_i$, and $\bm\beta^*\in\mathbb R^p$ is the unknown
parameter of interest. {\color{black}Here, $\{(\bm X_i, \epsilon_i)\}_{i=1}^{n}$ are independent and identically distributed (i.i.d.).}

We first revisit the penalized convoluted rank regression estimators in \cite{zhou2023sparse}. To estimate $\bm\beta^*$, \cite{hettmansperger2010robust} considered to minimize the following   loss function:
\begin{align}\label{loss}
    \min_{\bm\beta\in\mathbb{R}^p}
\frac{1}{n(n-1)}\sum_{i\neq j}^n|(Y_i-\bm X_i^\top\bm\beta)-(Y_j-\bm X_j^\top\bm\beta)|,
\end{align}
which is equivalent to minimize the Jaeckel's dispersion function in \cite{1972Jaeckel} with Wilcoxon scores \citep{wang2020tuning}.  
The $\bm\beta^*$ can be identified as the minimizer of the population version of the  loss function in (\ref{loss}), that is, $\bm\beta^*=\min_{\bm\beta\in\mathbb{R}^p}
\mathbb{E}[|(Y_i-\bm X_i^\top\bm\beta)-(Y_j-\bm X_j^\top\bm\beta)|].$ As shown by \cite{hettmansperger2010robust}, compared with the standard least squares method, the rank
regression estimator can have arbitrarily high relative
efficiency with heavy-tailed errors, while having
at least 86.4\% asymptotic relative efficiency under arbitrary symmetric error distribution with finite Fisher information. This robustness and high estimation efficiency makes the rank
regression estimator  a powerful choice. However the nonsmoothness of the rank loss brings heavy computation burden and hinders the popularity of rank estimator. 

To reduce the computation burden of rank
regression estimator, \cite{zhou2023sparse} innovatively considered the following convoluted rank loss function
\begin{align*}
   &L_h(u)=\int_{-\infty}^{\infty}|u-v|K_h(v)dv=u\int_{-u}^{u}K_h(v)dv-2\int_{-\infty}^{u}v K_h(v)dv, 
\end{align*}
in which $K_h(\cdot)=h^{-1}K(\cdot/h)$. 
Here $K(\cdot)$ is a kernel function and $h$ is a bandwidth. 
With the involvement of the {\color{black}smooth} kernel function {\color{black}satisfying Assumption (A2) in the following}, the convoluted rank loss function now becomes smooth and convex. This kind of convoluted smoothing technique has also been considered in \cite{fernandes2021smoothing}, \cite{he2023smoothed} and \cite{tan2022high} for quantile regression. A detailed discussion of the convoluted rank loss function can be found in  \cite{zhou2023sparse}. The population parameter is defined as the minimizer of the above convoluted rank loss function, that is, 
\begin{align*}
    &\bm\beta_h^\ast = \mathop{\arg\min}\limits_{\bm \beta\in \mathbb{R}^{p}}\E[L_h\left((Y_i-Y_j)-(\bm X_i-\bm X_j)^\top\bm\beta\right)].
\end{align*}
An interesting and important result is that for any $h>0$, $\bm\beta_h^\ast=\bm\beta^*$. That is, convoluted smoothing does
not incur any bias in the population level. This is different from convoluted quantile regression \citep{tan2022high}. As explained by \cite{zhou2023sparse}, the special form of rank regression makes the symmetry of the distributions of
$\epsilon_i-\epsilon_j$ and $\bm X_i-\bm X_j$ being a great benefit.

When the dimension is high, \cite{zhou2023sparse} proposed the following penalized convoluted rank regression estimators
\begin{align}
    &\widehat{\bm\beta}_h = \mathop{\arg\min}\limits_{\bm \beta\in \mathbb{R}^{p}}\frac{1}{n(n-1)}\sum_{i\neq j}^{n}L_h\left((Y_i-Y_j)-(\bm X_i-\bm X_j)^\top\bm\beta\right)+\sum_{j=1}^{p}p_{\lambda}(|\beta_j|).\label{CRR-solution}
\end{align}
Here, $p_{\lambda}(\cdot)$ is a penalty function with $\lambda$ being a tuning parameter. Under the assumption that the predictors are bounded, they obtained the estimation error bounds for the $\ell_1$ penalized convoluted rank regression. While under the minimal signal condition, they further established the oracle property of the folded concave penalized convoluted rank regression. However, the penalized convoluted rank regression estimators generally cannot be directly used to make inference of the parametric vector $\bm\beta^*$. In fact, as argued by \cite{zhang2008sparsity} and \cite{zhang2014confidence}, 
the penalized estimators with LASSO penalty are generally biased, while the minimal signal condition may be restrictive and may not be easy to verify. Further to simplify the theoretical analysis, \cite{zhou2023sparse} focused on the fixed design setting. Instead, we consider the random design setting, and we emphasize that this paper aims to develop inference procedures for convoluted rank regression. 

Now denote $\widehat \epsilon_i=Y_i-\bm X_i^\top\widehat{\bm\beta}_h$, and   \begin{align*}
    &\widehat{\mathbf J}_h = \frac{1}{n(n-1)}\sum_{i\neq j}^{n}L_h^{\prime\prime}\left(\widehat\epsilon_i-\widehat\epsilon_j\right)  (\bm X_i-\bm X_j)(\bm X_i-\bm X_j)^\top.
\end{align*}
Here, $\widehat{\mathbf J}_h$ is the sample version of the following matrix
\begin{align*}
&\mathbf J_h = \E \left\{L_h^{\prime\prime}(\epsilon_i-\epsilon_j) (\bm X_i-\bm X_j)(\bm X_i-\bm X_j)^\top\right\}=2\E [L_h^{\prime\prime}(\epsilon_i-\epsilon_j)] \mathbf\Sigma.
\end{align*}
Motivated by the debiasing technique developed recently (see for instance \cite{zhang2014confidence} and \cite{ning2017general}), we now consider the following debiased estimator 
\begin{align*}
    \widehat{\bm\beta}_h +\mathbf J_h ^{-1}\frac{1}{n(n-1)}\sum_{i\neq j}^{n}L_h^\prime\left(\widehat\epsilon_i-\widehat\epsilon_j\right)(\bm X_i-\bm X_j).
\end{align*}
It can be viewed as an one-step correction of the penalized estimator $\widehat{\bm\beta}_h$. In the above formula, $\mathbf J_h$ is unknown. Even if we can estimate $\mathbf J_h$ by $\widehat{\mathbf J}_h$, we cannot estimate $\mathbf J_h^{-1}$ directly by $\widehat{\mathbf J}_h^{-1}$, since when the dimension $p$ is larger than the sample size $n$, the matrix $\widehat{\mathbf J}_h$ may not be invertible. To this end, we can adopt the approach in \cite{yan2023confidence}. Denote the estimator of $\mathbf J_h^{-1}$ as $\widehat{\mathbf W}_h$. Then we obtain $\widehat{\mathbf W}_h$ via the following convex program: \begin{align}
  \min_{\mathbf W\in\mathbb R^{p\times p}}\left\|\mathbf W\right\|_{\ell_1},\quad {\mbox{s.t.}} \left\|   \mathbf W\widehat{\mathbf J}_h  -\mathbf I  \right\|_{\max}\leq \gamma_n.\label{optim_W_2}
\end{align}
Here $\gamma_n$ is a predetermined tuning parameter. 
{\color{black} From a practical viewpoint, there are alternative ways to estimate $\mathbf J_h^{-1}$, as it essentially boils down to estimating   $\mathbf\Sigma^{-1}$ and the quantity $\E L_h^{\prime\prime}(\epsilon_i-\epsilon_j)$. The former can be achieved using the same  techniques 
employed for the debiased LASSO \citep{van2014,javanmard2014confidence}, while the latter can be estimated by $\{n(n-1)\}^{-1}\sum_{i\neq j}^{n}L_h^{\prime\prime}\left(\widehat\epsilon_i-\widehat\epsilon_j\right)$.  }

{\color{black} Generally, $\widehat{\mathbf W}_h=\left(\widehat\omega_{ij}\right)_{i,j=1}^{p}$ is not symmetric since there
is no symmetry constraint on $\mathbf W$ in (\ref{optim_W_2}). To cope with this issue, one defines $\widehat{\mathbf W}_h^{\ast}=(\widehat \omega_{ij}^{\ast})_{i,j=1}^{p}$ with $\widehat \omega_{ij}^{\ast}=\widehat \omega_{ij}\mathbf{1}\left(\left|\widehat \omega_{ij}\right|\leq \left|\widehat \omega_{ji}\right|\right)+\widehat \omega_{ji}\mathbf{1}\left(\left|\widehat \omega_{ij}\right|> \left|\widehat \omega_{ji}\right|\right)$.
For technical convenience, we  note that  there is a little distinction
between (\ref{optim_W_2}) and the optimization problem (1) in \cite{cai2011constrained}, in which the constraint is imposed on $\left\|   \widehat{\mathbf J}_h \mathbf W -\mathbf I  \right\|_{\max}$.
}
Finally the debiased estimator is defined as:
\begin{align*}
     \widetilde{\bm\beta}_h=\widehat{\bm\beta}_h +\widehat{\mathbf W}_h\frac{1}{n(n-1)}\sum_{i\neq j}^{n}L_h^\prime\left(\widehat\epsilon_i-\widehat\epsilon_j\right)(\bm X_i-\bm X_j).
\end{align*}
Different from previous researches adopting debiasing technique, some term  in the  above is in the form of U-statistics. This brings new technical challenges for theoretical analysis. As an example, in \cite{zhou2023sparse}, they assumed the predictors are fixed and bounded to derive their theoretical results. In this paper, we first establish estimation error bounds of $\widehat{\bm\beta}_h$ under weaker conditions on the predictors. In particular, we only require the predictors to be sub-Gaussian, which is standard in high-dimensional analysis \citep{vershynin2018high}.

\par 
Some technical assumptions are introduced for theoretical development. 
\begin{itemize}
    \item[(A1)] Assume that $\bm X_i$ is a  $p$-dimensional sub-Gaussian random vector with $v_0:=\left\|\bm X_i\right\|_{\psi_2}$, and the eigenvalues of the covariance matrix $\mathbf\Sigma$ satisfy that $c_{\Sigma}\leq \lambda_{\min}(\mathbf\Sigma)\leq \lambda_{\max}(\mathbf\Sigma)\leq C_{\Sigma}$ for some positive constants $c_{\Sigma}$ and $C_{\Sigma}$.
    \item[(A2)] The kernel function $K(\cdot)$ satisfies that $K(v)\geq 0$, $K(-v)=K(v)$ for any $v\in\mathbb R$, $\sup_{v\in\mathbb R}K(v)<\infty$, $\sup_{v\in\mathbb R}K^\prime(v)<\infty$, and $\int_{\mathbb R} K(v)dv = 1$. Suppose that $\inf_{v\in[-1, 1]}K(v)=k_0>0$, and the bandwidth $h$ is bounded below by some positive constant $c_h$. 
     \item[(A3)] Let $g(\cdot)$ be the probability density function of $\epsilon_i-\epsilon_j$. We assume
$|g(x)-g(y)|\leq L_0|x-y|, \forall x, y\in\mathbb{R}$, for some constant $L_0> 0$. Further assume that $0<\underline{g}\leq g(0)\leq \overline{g}<\infty$. 
\item[(A4)] \begin{itemize}
    \item  [(i)]  The penalty  function $p_{\lambda}(\cdot)$ satisfies that $p_{\lambda}(0)=0$ and is symmetric around the zero $p_{\lambda}(t)=p_{\lambda}(-t)$ for all $t\in\mathbb R$.
    \item [(ii)] On the nonnegative real line, $p_{\lambda}(\cdot)$ is nondecreasing.
    \item[(iii)] For $t>0$, the function $t\to p_{\lambda}(t)/t$ is nonincreasing in $t$.
    \item[(iv)] The function $p_{\lambda}(\cdot)$ is differentiable for all $t\neq 0$ and has a right derivative at $t=0$ with $\lim_{t\to 0^{+}}p_{\lambda}^{\prime}(t)=\lambda a$ for some $a>0$.  
    \item[(v)] There exists $\mu\in(0,8c_{\Sigma}k_0\underline{g}/3)$ such that $p_{\lambda, \mu}(t)=:p_{\lambda}(t)+\mu t^2/2$ is convex. 
\end{itemize}  

\end{itemize}
The above conditions are very mild and commonly assumed. Assumption (A1) is a standard condition on the design $\bm X$ in the field of high-dimensional inference, see for instance \cite{van2014}, \cite{zhang2017simultaneous}  and \cite{yan2023confidence}. Besides, conditions on the kernel function in  Assumption (A2) are also assumed in \cite{zhou2023sparse} and \cite{tan2022high}. Similar to \cite{zhou2023sparse}, we do not need the bandwidth $h\to 0$   as $n\to\infty$ in the following theoretical analysis. 
Note that Assumption (A3) does not involve any moment assumption on the $\epsilon_i$'s, while \cite{zhang2017simultaneous} imposed an exponential-type tails condition on the error terms. 
Moreover, as discussed in \cite{loh2015regularized}, many commonly used penalty functions including the LASSO, SCAD \citep{fan2001variable}, and MCP \citep{zhang2010nearly} satisfy Assumption (A4).

\begin{remark}
     The condition $\inf_{v\in[-1, 1]}K(v)=k_0>0$ is only for theoretical and notational convenience.
If the kernel $K(\cdot)$ is compactly supported on $[-1, 1]$, we may rescale it to obtain $K_b(u)=K(u/b)/b$ for some $b > 1$. Then, $K_b(\cdot)$ is supported on $[-b, b]$ with $\inf_{v\in[-1, 1]}K_b(v) >0$. See also \cite{tan2022high} and \cite{yan2023confidence}. 
\end{remark}
Define $\|\bm\delta\|_{\bm\Sigma}=\|\bm\Sigma^{1/2}\bm\delta\|_2, \mathbb{B}_{\bm\Sigma}(r)=\{
\bm\delta\in\mathbb{R}^p: \|\bm\delta\|_{\bm\Sigma}\leq r\}$ and $\mathbb{C}_{\bm\Sigma}(l)=
\{
\bm\delta\in\mathbb{R}^p: \|\bm\delta\|_1\leq l \|\bm\delta\|_{\bm\Sigma}\}$. Further define the event 
$$\mathcal{A}(r,l,\kappa)=:\left\{\frac{\langle\nabla \mathcal{L}({\bm\beta}_h)-\nabla \mathcal{L}(\bm\beta_h^\ast), {\bm\beta}_h-\bm\beta_h^\ast\rangle\}}{\|\bm\beta_h-\bm\beta_h^\ast\|_{\bm\Sigma}^2}\geq \kappa, \,\,\mbox{for all}\,\,\bm\beta_h\in\bm\beta_h^\ast+\mathbb{B}_{\bm\Sigma}(r)\cap\mathbb{C}_{\bm\Sigma}(l)\right\},$$
{\color{black} where $\mathcal{L}({\bm\beta})=    
\{n(n-1)\}^{-1}\sum_{i\neq j}^{n}L_h\left((Y_i-Y_j)-(\bm X_i-\bm X_j)^\top\bm\beta\right)$.}
The following Proposition \ref{localconvexity} indicates that, when $(r,l,\kappa)$  are chosen appropriately,  the event $\mathcal{A}(r,l,\kappa)$ occurs with a high probability. 
\begin{proposition}\label{localconvexity}
Under Assumptions (A1)-(A3),  and further  let 
$$\kappa=2k_0\underline{g},\,\,\,l/r\sqrt{\log p/n}=\o(1),\,\,\,\mbox{and}\,\,\, 64c_{\Sigma}^{-1}v_0^2 r\leq h\leq \underline{g}/(2L_0),$$
we have
$\mathbb{P}\{\mathcal{A}(r,l,\kappa)\}\rightarrow 1$, as $n\to\infty$.  
\end{proposition}
The above proposition establishes the local restricted strong convexity for convoluted rank regression in high
dimensions. With  Proposition \ref{localconvexity}, we can obtain the following result, which establishes the estimation error bound of $\widehat{\bm\beta}_h$ under sub-Gaussion assumptions of the predictors. 
  Due to the potential nonconvexity of the penalty function, we include a side constraint that $\left\|\bm\beta\right\|_1\leq R$ in the optimization problem (\ref{CRR-solution}), where $R$ is allowed to tend to infinity. 
Now let $\mathcal{S}=\{j: \beta^*_j\neq 0\}$ and $s=|\mathcal{S}|$.

\begin{theorem}\label{estimationerror-theorem}
 
Suppose that   $\log(p)=\O(\sqrt{n})$, and there exist $r_n=\o(1)$ and $\tau_n\to \infty$ such that 
 $\tau_n R^2\log p =\o(nr_n^2)$  and $\tau_n  r_n^2\rightarrow\infty$. Under Assumptions (A1)-(A4), with  the choice of $\lambda\asymp \sqrt{\log(p)/n}$ and $64c_{\Sigma}^{-1}C_{\Sigma}^{1/2}v_0^2 r_n\leq h\leq \underline{g}/(2L_0)$,  
the penalized convoluted rank regression estimator $\widehat{\bm\beta}_h$  satisfies   
    the following estimation error bound 
    \begin{align}\label{error-bound}
        \left\|\widehat{\bm\beta}_h-\bm\beta_h^\ast\right\|_1=\O_p\left(s\sqrt{\frac{\log(p)}{n}}  \right),\,\,
         \left\|\widehat{\bm\beta}_h-\bm\beta_h^\ast\right\|_2=\O_p\left(\sqrt{\frac{s\log(p)}{n}}  \right).
    \end{align}

Specially, when the penalty function $p_{\lambda}(t)=\lambda|t|$,  the constraint $\left\|\bm\beta\right\|_1\leq R$  becomes unnecessary. Under Assumptions (A1)-(A3) and $\log(p)=\O(\sqrt{n})$, with $\lambda\asymp \sqrt{\log(p)/n}$ and 
$\sqrt{s\log(p)\log(n)/n}\lesssim h\leq \underline{g}/(2L_0)$, the results in (\ref{error-bound}) hold for LASSO penalty.

\end{theorem}
The above theorem is a generic result. Same as  \cite{loh2015regularized} and \cite{wang2024analysis}, we treat many commonly used penalty functions in a unified way. Besides, the above result also extends the results in \cite{zhou2023sparse} in two aspects. {\color{black}Firstly \cite{zhou2023sparse} assumed that the predictors are fixed design and bounded, while in our paper we consider random design setting and assume that the predictors are sub-Gaussian, which is milder than the boundedness assumption, and is commonly assumed in high-dimensional setting. Secondly, we derive the estimation error bounds of penalized estimators with many penalty functions including LASSO, SCAD and MCP in a unified approach. For nonconvex penalty functions, our results show that all local minima within a feasible region have desirable error bounds. As noted by  \cite{loh2015regularized} and \cite{wang2024analysis}, the global minimizers for nonconvex objective functions are not always numerically
 obtained or verifiable in practice. Thus these new results are important.
 
 }



To further delve into the asymptotic behavior of the proposed debiased estimator $\widetilde{\bm\beta}_h$, the following Theorem \ref{bahadur-theorem} concerns the Bahadur representation of $\widetilde{\bm\beta}_h-\bm\beta_h^\ast$. 
\begin{itemize}
    \item[(B1)] Assume that $\mathbf J_h^{-1}= \left({\bm\omega}_{1},\ldots,{\bm\omega}_{p}\right)^\top$ is row-wisely sparse, i.e., $\max_{1\leq i\leq p}\sum_{k=1}^{p}\left|\omega_{ik}\right|^q<C_\omega$ for some  $q\in(0,1)$ and positive constant $C_\omega$. The tuning parameter $\gamma_n\asymp sn^{-1/2}\log^2(p)$. 
\end{itemize}
This assumption  requires $\mathbf J_h^{-1}$
to be sparse in the sense of $\ell_q$-norm of matrix row
space, which   is crucial for estimation of the inverse Hessian matrix. Similar conditions were widely   considered in the literature, see \cite{Bickel2008}, \cite{cai2011constrained}, and \cite{Cai2016}. 

\begin{theorem}\label{bahadur-theorem}
Suppose that $\widehat{\bm \beta}_h$ satisfies the estimation error bound in Theorem \ref{estimationerror-theorem}.   
Under Assumptions (A1),(A2),(B1) and $\log(p)=\O(\sqrt n)$,  the debiased estimator $\widetilde{\bm\beta}_h$ enjoys the following Bahadur representation:  \begin{align*}
  &\left\| 
\sqrt{n}\left(\widetilde{\bm\beta}_h-\bm\beta_h^\ast\right)-\sqrt{n}\mathbf J_h^{-1} \frac{1}{n(n-1)}\sum_{i\neq j}^{n}L_h^\prime\left(\epsilon_i- \epsilon_j\right)(\bm X_i-\bm X_j) \right\|_{\infty}
 \\&=\O_p\left(\left\{s\log^{(5-4q)/(2-2q)}(p)/\sqrt n\right\}^{1-q}+ s^2\log^{5/2}(p)/\sqrt n\right).
\end{align*}
\end{theorem}
Theorem \ref{bahadur-theorem} establishes the uniform Bahadur representation  for the debiased convoluted rank regression estimator, 
which suggests that $\sqrt{n}(\widetilde{\bm\beta}_h-\bm\beta_h^\ast)$ can be  expressed as a high-dimensional U-statistic up to some negligible terms. With its help, the asymptotic distribution of the estimators can be derived in conjunction
with some Gaussian approximation results in Section \ref{sec3}.

\section{Simultaneous inference}\label{sec3}
In the following, we develop simultaneous confidence intervals for  $\{ \beta_k^\ast  \}_{k\in\mathcal G}$. Here  $\mathcal G\subseteq  \{1,\ldots,p\}$ is an index set of interest. Simultaneous inference is a powerful tool in genetics, as a gene pathway, consisting of many genes for the same biological function, should be investigated simultaneously. To construct simultaneous confidence intervals, we need to focuse on the distribution of uniform deviations rather than coordinate-wise asymptotic distributions. For notation convenience, define $\left\|\bm a\right\|_{\mathcal G}=\max_{k\in \mathcal G}|a_k|$ for some vector $\bm a\in\mathbb R^p$.

\subsection{High-dimensional Gaussian approximation}
\begin{itemize}
    \item[(C1)] Let $f(\cdot)$ be the probability density function of $\epsilon_i$. Assume that $f(\cdot)$ is continuous, and there exists a positive constant   $\delta_1$ such that $\inf_{v\in[-\delta_1,\delta_1]}f(v)>0$.
\end{itemize}

\begin{theorem}\label{theorem3}
Suppose that $\widehat{\bm \beta}_h$ satisfies the estimation error bound in Theorem \ref{estimationerror-theorem},  {\color{black}$\log^5(np)/n^{1-\iota}=\O(1)$} for some  $\iota\in(0,1)$, 
$s\log^{(3-2q)/(1-q)}(p)/\sqrt n=\o(1)$ and $s^2\log^{3}(p)/\sqrt n=\o(1)$.
    Under Assumptions (A1), (A2), (B1), (C1), there exists   a $p$-dimensional Gaussian random vector ${\bm Z}_n$   with mean zero and covariance \begin{align}\label{asymp-variance}
      {\mathbf \Sigma}_Z=\E \left( {\bm Z}_n  {\bm Z}_n^\top \right)  = \frac{  \E \left[\E \left\{L_h^{\prime}(\epsilon_i-\epsilon_j)|\epsilon_i\right\}\right]^2}{\left\{\E L_h^{\prime\prime}(\epsilon_i-\epsilon_j)\right\}^2} \mathbf\Sigma^{-1}, 
   \end{align} such that  
    \begin{align*}
    \sup_{t\in\mathbb R}\left|\P\left(\sqrt n\left\|\widetilde{\bm\beta}_h-\bm\beta_h^\ast\right\|_{\mathcal G}\leq t\right)-\P\left(\left\|\bm Z_n\right\|_{\mathcal G}\leq t\right)\right|=\o(1).
\end{align*}
\end{theorem}
The above result establishes the high-dimensional Gaussian approximation result of $\sqrt n(\widetilde{\bm\beta}_h-\bm\beta_h^\ast)$, indicating that the distribution of the max-norm of $\sqrt{n}(\widetilde{\bm\beta}_h-\bm\beta_h^\ast)$ is asymptotically equal to the maximum of a Gaussian random vector  with covariance matrix $\mathbf{\Sigma}_Z$. 
 \par By comparing the asymptotic  variance of $\widetilde{\bm\beta}_h$ with the well-known desparsifying  estimator $\widetilde{\bm\beta}$ considered in the literature,  see \cite{van2014} and \cite{zhang2017simultaneous} for example, we can show the  asymptotic relative efficiency of $\widetilde{\bm\beta}_h$ with respect to $\widetilde{\bm\beta}$. 
 {\color{black} We observe that the asymptotic  covariance matrices of 
$\widetilde{\bm\beta}_h$ and $\widetilde{\bm\beta}$
  differ only by a scalar factor, which is referred to as the asymptotic relative efficiency (ARE)
of $\widetilde{\bm\beta}_h$ with respect to   $\widetilde{\bm\beta}$. This is also adopted by \cite{zhou2023sparse}. }
\begin{corollary}\label{corollary1}
The asymptotic relative efficiency (ARE)
of $\widetilde{\bm\beta}_h$ with respect to the desparsifying  estimator $\widetilde{\bm\beta}$ considered in \cite{van2014} and \cite{zhang2017simultaneous} is given as follows
    \begin{align*}
{\mbox{ARE}}\left(\widetilde{\bm\beta}_h ,\widetilde{\bm\beta}\right)=\frac{\E\epsilon_i^2 \left\{\E L_h^{\prime\prime}(\epsilon_i-\epsilon_j)\right\}^2}{  \E \left[\E \left\{L_h^{\prime}(\epsilon_i-\epsilon_j)|\epsilon_i\right\}\right]^2}.
    \end{align*}
    Moreover, we have that  $$\lim_{h\to 0_+}{\mbox{ARE}}\left(\widetilde{\bm\beta}_h ,\widetilde{\bm\beta}\right)=12\left\{\int f^2(x)dx\right\}^2\E\epsilon_i^2. $$
\end{corollary}
When $h$ is away from $0$, it is hard to give a simple expression  for  ${\mbox{ARE}}\left(\widetilde{\bm\beta}_h ,\widetilde{\bm\beta}\right)$, but given the kernel function, the bandwidth $h$ and the error distribution, we could directly compute the value of ${\mbox{ARE}}\left(\widetilde{\bm\beta}_h ,\widetilde{\bm\beta}\right)$  using some numerical integration
tool. 
Consider the 
Epanechnikov kernel $K(u) = 3/4(1 - u^2)\mathbf{1}_{\{ -1 \leq u \leq 1\}}$ with the bandwidth $h=1$  for instance. 
When the error distribution is standard normal, the value of  ${\mbox{ARE}}\left(\widetilde{\bm\beta}_h ,\widetilde{\bm\beta}\right)$ is  $0.96$; 
when the error follows a standard Cauchy distribution, the value is $\infty$,  since the second moment of $\epsilon_i$ does not exist. {\color{black}  Numerical results are presented in Table S.6 in the supplemental material to evaluate the value of ${\mbox{ARE}}\left(\widetilde{\bm\beta}_h ,\widetilde{\bm\beta}\right)$ under different distributions of the error term, which align with our theoretical findings. }

{\color{black}
In the following, we make some theoretical comparison between our proposed debiased CRR estimator and other recently developed debiased robust estimators in terms of ARE. Firstly, the following corollary compares the ARE between our proposed debiased CRR estimator $\widetilde{\bm\beta}_h$ and the debiased Huber loss estimator $\bar{\bm\beta}_{ \tau}$ proposed in \cite{loh2021scale}. 
\begin{corollary}
Let $\Psi_{\tau}(u)=u\bm{1}_{\{|u|\leq \tau\}}+\tau\bm{1}_{\{|u|>\tau\}}$. 
    The ARE between our proposed debiased CRR estimator $\widetilde{\bm\beta}_h$ and the debiased Huber loss estimator $\bar{\bm\beta}_{\tau}$  is given by \begin{align*}
    &{\mbox{ARE}}\left(\widetilde{\bm\beta}_h ,\bar{\bm\beta}_{\tau}\right)=\frac{\E\Psi_{\tau}^2(\epsilon_i)  \left\{\E L_h^{\prime\prime}(\epsilon_i-\epsilon_j)\right\}^2}{  \left\{\E\Psi^\prime(\epsilon_i)\right\}^2  \E \left[\E \left\{L_h^{\prime}(\epsilon_i-\epsilon_j)|\epsilon_i\right\}\right]^2 },\\
    &\lim_{h\to0_+}{\mbox{ARE}}\left(\widetilde{\bm\beta}_h ,\bar{\bm\beta}_{\tau}\right)=12\left\{\int f^2(x)dx\right\}^2\frac{\E\Psi_{\tau}^2(\epsilon_i)  }{  \left\{\E\Psi^\prime(\epsilon_i)\right\}^2    }.
\end{align*} 

\end{corollary}
Let $h\to 0$ and $\tau=3$ for instance. 
When the error distribution is standard normal, the value of  $\lim_{h\to0_+}{\mbox{ARE}}\left(\widetilde{\bm\beta}_h ,\bar{\bm\beta}_{\tau}\right)$ is  $0.96$; 
when the error follows a standard Cauchy distribution, the value is $1.42$. 
}

{\color{black}We also compare the ARE of our proposed debiased estimator $\widetilde{\bm\beta}_h$ with the debiased convoluted quantile regression in \cite{yan2023confidence} and the debiased composite quantile regression in \cite{zhao2014general}. 
It is noted that the theory in \cite{yan2023confidence} and \cite{zhao2014general} both require $h\to 0$, though our method can work   without this requirement.

\begin{corollary}
The ARE between our proposed debiased CRR estimator $\widetilde{\bm\beta}_h$ and the convoluted quantile estimator $\breve{\bm\beta}_{h}$ proposed in \cite{yan2023confidence} with quantile level $\tau=0.5$ is given by \begin{align*}
    {\mbox{ARE}}\left(\widetilde{\bm\beta}_h ,\breve{\bm\beta}_{h}\right)=\frac{ \E\left\{ \int_{\infty}^{-\epsilon_i}K_h(t)dt-1/2 \right\}^2 \left\{\E L_h^{\prime\prime}(\epsilon_i-\epsilon_j)\right\}^2}{  \left\{\E K_h(-\epsilon_i)\right\}^2  \E \left[\E \left\{L_h^{\prime}(\epsilon_i-\epsilon_j)|\epsilon_i\right\}\right]^2 }.
\end{align*} 
Letting $h\to 0$, we have   that \begin{align*}
    \lim_{h\to 0_+}{\mbox{ARE}}\left(\widetilde{\bm\beta}_h ,\breve{\bm\beta}_{h}\right) = 3\left\{\int f^2(x)dx\right\}^2/ f^2(0).
\end{align*}
\end{corollary}

By letting $h\to 0$,  when $\epsilon$ follows the standard normal distribution, the asymptotic relative efficiency is calculated by $1.5$; when $\epsilon$ follows the standard Cauchy distribution, the asymptotic relative efficiency is calculated by $0.75$. 
\par The following corollary states that the debiased composite quantile estimator  $\check{\bm\beta}_{  K^*}$ proposed in \cite{zhao2014general} and our proposed $\widetilde{\bm\beta}_h$ have the same asymptotic variance when $K^*\to\infty$
and $h\to 0$.
\begin{corollary}

The ARE between our proposed debiased CRR estimator $\widetilde{\bm\beta}_h$ and the debiased composite quantile estimator  $\check{\bm\beta}_{K^*}$ proposed in \cite{zhao2014general} with $K^*\to\infty$  is given by \begin{align*}
    &\lim_{K^*\to\infty}{\mbox{ARE}}\left(\widetilde{\bm\beta}_h ,\check{\bm\beta}_{ K^*}\right)=\frac{  \left\{\E L_h^{\prime\prime}(\epsilon_i-\epsilon_j)\right\}^2}{ 12\left\{\int f^2(x)dx\right\}^2  \E \left[\E \left\{L_h^{\prime}(\epsilon_i-\epsilon_j)|\epsilon_i\right\}\right]^2 },
    \\
    &\lim_{h\to 0_+} \lim_{K^*\to\infty}{\mbox{ARE}}\left(\widetilde{\bm\beta}_h ,\check{\bm\beta}_{K^*}\right)=1.
\end{align*}
\end{corollary}

The above kind of equivalence of the asymptotic variance matrices has also been noticed by \cite{wang2020tuning}. When $K^*$ is not too large or the $h$ is not close to zero, these two estimators may not have the same asymptotic variance matrix. In practice, the debiased composite quantile estimator $\check{\bm\beta}_{  K^*}$ proposed in \cite{zhao2014general} requires to estimate the density function of the error term at $K^*$ points. From our simulation results, the performance of debiased composite quantile estimator $\check{\bm\beta}_{  K^*}$ largely depends on the estimation accuracy of the estimated density function.
}

\subsection{A novel bootstrap procedure}
To proceed, in order to construct simultaneous confidence regions for $\{ \beta_k^\ast  \}_{k\in\mathcal G}$, it suffices to obtain the upper-$\alpha$ quantile of the $\mathcal G$-norm of $\bm Z_n$.  However,  $\bm Z_n$ is a high-dimensional Gaussian random vector with unknown covariance matrix ${\mathbf \Sigma}_Z\neq {\mathbf I}$, estimating  which involves more nuisance parameters. Besides, even if the covariance matrix ${\mathbf \Sigma}_Z$ can be estimated accurately,
simulating the distribution of $\left\|\bm Z_n\right\|_{\mathcal G}$ by generating resamples from the high-dimensional Gaussian distribution is time consuming and computation intensive. Alternatively, we may approximate the asymptotic distribution of $\left\|\bm Z_n\right\|_{\mathcal G}$ by extreme value distribution \citep{ma2021global}. However, this generally requires further restriction on the covariance matrix ${\mathbf \Sigma}_Z$ and the convergence rate is generally slow. Motivated by our Bahadur representation in Theorem \ref{bahadur-theorem}, we may consider the bootstrapped version of 
$$\left\|\sqrt{n}\mathbf J_h^{-1} \frac{1}{n(n-1)}\sum_{i\neq j}^{n}L_h^\prime\left(\epsilon_i- \epsilon_j\right)(\bm X_i-\bm X_j) \right\|_{\mathcal G}.$$
Note that the above expression is a U-statistic. Recently \cite{chen2017} developed the Gaussian and boostrap approximations for high-dimensional U-statistics. However, different from \cite{chen2017}, $\mathbf J_h^{-1}L_h^\prime\left(\epsilon_i- \epsilon_j\right)(\bm X_i-\bm X_j)$ depends on unknown quantities $\epsilon_i$'s and $\mathbf J_h^{-1}$, and thus their theoretical results cannot be directly adopted here. 

We now consider the following multiplier bootstrap procedure. Let $e_1,\cdots,e_n$ be iid $N(0, 1)$ random variables that are also independent of the data $\mathcal{D}=\{\bm X_i,Y_i\}_{i=1}^{n}$. Consider
$$\bm T_{boot}=\sqrt{n}\widehat{\mathbf W}_h  \frac{1}{n(n-1)}\sum_{i\neq j}^{n}L_h^\prime\left(\widehat\epsilon_i- \widehat\epsilon_j\right)(\bm X_i-\bm X_j)(e_i+e_j).$$
In  $\bm T_{boot}$, we make a slight modification of the multiplier bootstrap procedure in \cite{chen2017}. Actually if we directly follow \cite{chen2017}, we may consider the following form
$$\sqrt{n}\widehat{\mathbf W}_h  \frac{1}{n(n-1)}\sum_{i\neq j}^{n}L_h^\prime\left(\widehat\epsilon_i- \widehat\epsilon_j\right)(\bm X_i-\bm X_j)\{(e_i+1)(e_j+1)-1\}.$$
We find that the additional term with $e_ie_j$ is not necessary and is removed here. It is noted that no high-dimensional random vectors are resampled here, instead we only require to generate one-dimensional $e_i$'s, resulting in a low computational burden. The following theorem establishes the validity of the proposed bootstrap procedure.

\begin{theorem}\label{theorem4}
Suppose that $\widehat{\bm \beta}_h$ satisfies the estimation error bound in Theorem \ref{estimationerror-theorem}.
    Under Assumptions (A1),(A2),(B1),(C1), and {\color{black}$sn^{-1/2}\{\log(p)\}^{(3-2q)/(1-q)}\{\log\log(p)\}^{1/(1-q)} =\o(1)$,} 
    we have
$$\sup_{t\in\mathbb R}\left|\P\left(\left\|\bm T_{boot}\right\|_{\mathcal G}\leq t|\mathcal{D}\right)-\P\left(\left\|\bm Z_n\right\|_{\mathcal G}\leq t\right)\right|\xrightarrow{\P}0.$$
\end{theorem}

Denoted by $Q_{1-\alpha}$ the upper-$\alpha$ quantile of $\left\|  \bm T_{boot}\right\|_{\mathcal G}$. From Theorems \ref{theorem3} and \ref{theorem4}, one constructs asymptotically correct simultaneous confidence intervals of $\{\beta_{h,k}^{\ast}\}_{k\in\mathcal G}$ as follows \begin{align*}
    \left(\widetilde{ \beta}_{h,k}-n^{-1/2}Q_{1-\alpha},\widetilde{ \beta}_{h,k}+n^{-1/2}Q_{1-\alpha}\right),\,\,\,\forall k\in\mathcal G.
\end{align*}
From the above formula, it is clear that the length of each confidence interval is fixed for any $k\in\mathcal G$. In order to obtain simultaneous confidence intervals with varying length and avoid estimating the nuisance quantity in the numerator of (\ref{asymp-variance}), we consider the following studentized version  $\sqrt{n}\widehat\omega^{-1/2}_{kk} (\widetilde\beta_{h,k}-\beta_{h,k}^\ast )$, in which $\widehat\omega_{kk}$ is the $(k,k)$-element of the matrix $\widehat{\mathbf W}_h$. Note that $\widehat\omega^{1/2}_{kk}$ is not the estimated standard deviation of $\sqrt{n} (\widetilde\beta_{h,k}-\beta_{h,k}^\ast )$, but differs only by a  nuisance constant factor, which does not depend  on $k$.  Denote by 
    $\widehat{\mathbf S} ={\mbox{diag}}\left(\widehat{\mathbf W}_h\right)$ and $ {\mathbf S} ={\mbox{diag}}\left(  \mathbf J_h^{-1}\right)$.
Define   \begin{align*}
    \bm T^{\star}_{boot}=\sqrt{n}\widehat{\mathbf S}^{-1/2}\widehat{\mathbf W}_h  \frac{1}{n(n-1)}\sum_{i\neq j}^{n}L_h^\prime\left(\widehat\epsilon_i- \widehat\epsilon_j\right)(\bm X_i-\bm X_j)(e_i+e_j).
\end{align*}
\begin{theorem}\label{theorem5}
    Under the conditions of Theorem \ref{theorem3}, there exists a $p$-dimensional Gaussian random vector $\bm Z_n^\star$ with mean zero and covariance 
    \begin{align*}
    \E\left(\bm Z^{\star}_{n}{\bm Z^{\star}_{n}}^\top\right) = \frac{  \E \left[\E \left\{L_h^{\prime}(\epsilon_i-\epsilon_j)|\epsilon_i\right\}\right]^2}{\left\{\E L_h^{\prime\prime}(\epsilon_i-\epsilon_j)\right\}^2} \mathbf S^{-1/2}\mathbf\Sigma^{-1}\mathbf S^{-1/2},
\end{align*}
such that 
\begin{align*}
    \sup_{t\in\mathbb R}\left|\P\left(\sqrt n\left\|\widehat{\mathbf S}^{-1/2}\left(\widetilde{\bm\beta}_h-\bm\beta_h^\ast\right)\right\|_{\mathcal G}\leq t\right)-\P\left(\left\|\bm Z^{\star}_{n}\right\|_{\mathcal G}\leq t\right)\right|=\o(1).
\end{align*}
Furthermore, under the conditions of Theorem \ref{theorem4}, we derive that 
\begin{align*}
    \sup_{t\in\mathbb R}\left|\P\left(\left\|\bm T^{\star}_{boot}\right\|_{\mathcal G}\leq t|\mathcal{D}\right)-\P\left(\left\|\bm Z^{\star}_n\right\|_{\mathcal G}\leq t\right)\right|=\o(1).
\end{align*}
\end{theorem}


Denoted by $Q_{1-\alpha}^{\star}$ the upper-$\alpha$ quantile of $\left\| \bm T_{boot}^{\star}\right\|_{\mathcal G}$. Inspired by Theorem \ref{theorem5}, one constructs the simultaneous confidence intervals for $\{\beta_{h,k}^{\ast}\}_{k\in\mathcal G}$ with varying width as follows \begin{align*}
    \left(\widetilde{ \beta}_{h,k}-\widehat\omega_{kk}^{1/2}n^{-1/2}Q_{1-\alpha}^{\star},\widetilde{ \beta}_{h,k}+\widehat\omega_{kk}^{1/2}n^{-1/2}Q_{1-\alpha}^{\star}\right),\,\,\,\forall k\in\mathcal G.
\end{align*} 
We summarize the detailed steps for implementation in Algorithm \ref{alg2}. 

\begin{algorithm}\label{alg2}
	\renewcommand{\algorithmicrequire}{\textbf{Input:}}
	\renewcommand{\algorithmicensure}{\textbf{Output:}}
	\caption{Bootstrap procedures for simultaneous confidence intervals  with varying width. } 
	\label{alg2} 
	\begin{algorithmic}[1]
 \REQUIRE  Data $\{\bm X_i,Y_i\}_{i=1}^{n}$, the kernel $K(\cdot)$,  the bangwidth $h$, the debiased estimator $\widetilde{\bm\beta}_h$, the estimated Hessian matrix $\widehat{\mathbf W}_h$,   the residuals $\{\widehat{\epsilon}_i\}_{i=1}^{n}$, the index set $\mathcal G$, and the siginificance level $\alpha$.
 \STATE Compute $\widehat{\mathbf S}={\mbox{diag}}\left(\widehat{\mathbf W}_h\right)={\mbox{diag}}\left(\widehat\omega_{11},\ldots,\widehat\omega_{pp}\right)$. 
\FOR{$b=1,2,\ldots,B$}
\STATE Generate standard normal random variables $\{e_{bi}\}_{i=1}^{n}$. 
\STATE Compute $\bm T_{boot,b}^{\star}=\sqrt{n}\widehat{\mathbf S}^{-1/2}\widehat{\mathbf W}_h  \{n(n-1)\}^{-1}\sum_{i\neq j}^{n}L_h^\prime\left(\widehat\epsilon_i- \widehat\epsilon_j\right)(\bm X_i-\bm X_j)(e_{bi}+e_{bj})$.
\ENDFOR
\STATE Compute the upper-$\alpha$ quantile of $\left\{\left\| \bm T_{boot,b}^{\star}\right\|_{\mathcal G}\right\}_{b=1}^{B}$, denoted by $\widehat Q_{1-\alpha}^{\star}$. 
\ENSURE The simultaneous confidence intervals $\widetilde{\beta}_{h,k}\pm \widehat{\omega}_{kk}^{1/2}n^{-1/2} \widehat Q_{1-\alpha}^{\star}$ for $k\in\mathcal G$.
\end{algorithmic} 
\end{algorithm}

\section{Numerical simulation}\label{sec4}
In this section, we conduct  numerical simulations to evaluate the finite sample performance of our proposed simultaneous inference procedures.  {\color{black}For comparison, we also implement the simultaneous confidence intervals proposed in \cite{zhang2017simultaneous}, denoted by ZC; the robust inference procedures proposed in \cite{yan2023confidence}, denoted by YWZ; the robust confidence regions proposed in \cite{zhao2014general}  with $K^*=9$ as suggested, denoted by ZKL ($f(\cdot)$ is unknown) and ZKL$^\ast $ ($f(\cdot)$ is known); the confidence regions for high-dimensional sparse median regression model proposed in \cite{belloni2015uniform}, denoted by BCK. As suggested by a reviewer, although \cite{yan2023confidence} has not developed a method to construct simultaneous confidence intervals, we can achieve this goal based on their Theorems 3 and 4 and Theorem 2.6 of \cite{belloni2018high}. Additionally, we note that our method can be adopted to constructing confidence intervals for an individual coefficient or extend to  significance testing. We have also compared our method with the robust post-selection inference method based on Huber loss in \cite{han2022robust}, the  robust residual-based inference procedures in \cite{wang2021robust} and \cite{guo2023invariance}. Due to the space limitation, additional simulations are presented in the supplementary materials.
}

In our numerical studies, the recommended technique for selecting the tuning parameter $\lambda$ in the LASSO penalty is through $10$-fold cross-validation (CV); see \cite{10.1214/009053607000000127}, \cite{hastie2009elements} for instance.  For folded concave penalty functions, for example the SCAD, we suggest use the high-dimensional BIC criteria developed in \cite{wang2013calibrating} and \cite{zhou2023sparse} for the choice of $\lambda$ and the local linear approximation algorithm in \cite{zou2008one} and \cite{zhou2023sparse} for efficient computation. 
For the   estimation of $\widehat{\mathbf W}_h$, we suggest   
the R package {\textsf{flare}} (\cite{rpackage}) with the default choice of the tuning parameter $\gamma_n$. 

The  Epanechnikov kernel is adopted as the density
convolution kernel, $K(u)=3/4(1-u^2)\mathbf{1}_{\{-1\leq u\leq 1\}}$, where
$\mathbf{1}_{\{\cdot\}}$ is the indicator function, and the loss function is defined by \begin{align*}
    L_h(u)=\left\{\begin{matrix}
 u & u\geq h,\\
3u^2h^{-1}/4-u^4h^{-3}/8+3h/8  & -h<u<h,\\
 -u & u\leq -h.
\end{matrix}\right.
\end{align*}
As suggested in \cite{zhou2023sparse}, we fix $h=1$. Though CV can be used to select 
$h$ as an additional tuning parameter, we opt not to do this to avoid the possibility that the improvement of convoluted rank regression over least square regression may come from this additional tuning.
 {\color{black} We also compared   different choices of $h=0.1,0.25,0.5,1,2$  and several other kernel functions (the Gaussian kernel, the biweight kernel, and the cosine kernel), and the overall results remain consistent. These additional simulation results can be found in Tables S.1-S.4 in the supplementary materials, which demonstrate the robustness of our method to hyperparameter selection.}

\par Consider the following model: 
    $Y_i = \bm X_i^\top \bm\beta+\epsilon_i$.
The predictors $\bm X_i$'s are independently generated from a multivariate normal distribution $N(0,\mathbf\Sigma)$, and $\epsilon_i$'s are i.i.d. random variables.  In the following, we set $\bm\beta = (\sqrt 3,\sqrt 3,\sqrt 3,0,\ldots,0)^\top\in\mathbb R^p$, in which  only the first three components are nonzeros. {\color{black}Besides, to evaluate the performance under different combination of the sparsity level $s$ and the signal strength, we further consider different settings of $\bm\beta$, and the corresponding results are illustrated in Figures S.1-S.3 in the supplemental material.} Different examples are considered to assess the performance of the proposed methods. 
\begin{itemize}
    \item[(i)] (Feature correlations) We consider two  scenarios
for the correlation structure of $\bm X$: (a)  $\mathbf\Sigma$ is Toeplitz with $\Sigma_{ij}=0.5^{|i-j|}$;  (b) $\mathbf\Sigma$ is banded, that is, $\Sigma_{ii}=1$ for $i=1,\ldots,p$ and $\Sigma_{ij}=0.48\mathbf{1}_{\{|i-j|=1\}}$ for $i\neq j$.
\item[(ii)] (Error distributions)  The errors $\epsilon_i$'s are    generated from the following distributions: (a) the standard normal distribution $N(0,1)$; (b) the mixture normal distribution $0.95 N(0,1)+0.05N(0,100^2)$, i.e. $\P(\epsilon\overset{d}{=}  N(0,1))=0.95$ and $\P(\epsilon\overset{d}{=}  N(0,100^2))=0.05$; (c) the standard Cauchy distribution. 
\item[(iii)] (Dimension of covariates) We fix the sample size $n = 100$ and use
the dimensions $p = 50,100,200$.
\item[(iv)] (The number of hypothesized predictors) We are interested in making simultaneous inference for the coefficients in certain index set $\mathcal G$. Different choices of $\mathcal G$ are taken into consideration: $\{1,2,3,4,5\}$, $\{1,2,\ldots,[p/5]\}$, and $\{1,2,\ldots,p\}$.
\item[(v)] (The penalty function for estimation) We consider both the LASSO penalty $p_{\lambda}(|t|)=\lambda|t|$, denoted by CRR-LASSO, and the  SCAD penalty proposed in \cite{fan2001variable}, \begin{align*}
    p_{\lambda}(|t|)=\lambda|t|\mathbf{1}_{\{0\leq |t|< \lambda \}}+\frac{a\lambda|t|-(t^2+\lambda^2)/2}{a-1}\mathbf{1}_{\{\lambda \leq |t|\leq  a\lambda \}}+\frac{(a+1)\lambda^2}{2}\mathbf{1}_{\{|t|>a\lambda\}},
\end{align*} 
with default choice of parameter $a=3.7$, denoted by CRR-SCAD. 
\end{itemize}

 The empirical coverage rates and the interval
widths are computed based on $500$ simulation runs, and the multiplier bootstrap replications $B=500$. {\color{black} As suggested by one anonymous referee, we also tried different choices of $B$ (e.g., 100, 200, 1000), the results are similar, provided in Table S.5 in the supplemental material. Additionally, to demonstrate the method's practical impact, we present
 the computing time of the proposed methods under various parameter configurations in Table S.7 in the supplemental material. When $n$ becomes too large, since the loss
 function in our method relies on second-order U-statistics (with respect to $n$),
 we develop a Divide and Conquer algorithm for computational efficiency. Kindly
 see pages 12-14 of the supplementary materials.}

From Tables \ref{table1}-\ref{table3}, we have the following findings. Firstly, considering the empirical coverage rates our method outperforms other
methods in  most cases, and the simulation results in Tables \ref{table2}-\ref{table3} where the errors follow the mixture normal distribution and the standard Cauchy  distribution provide a strong
evidence that our proposed  procedure is sufficiently robust and hardly affected by the heavy-tailed or outliers. Secondly, the average lengths of ZC are notably wide in Tables \ref{table2}-\ref{table3}, and the empirical coverage rates of ZC  fail to reach the nominal coverage rates when the dimension of coefficients is high or the design of covariates is highly correlated in Table \ref{table1}. {\color{black}Thirdly, the average length of the simultaneous  confidence intervals in YWZ is wider than that of our method in various scenarios.  Fourthly, in practical applications, the density function of the error term is unknown. BCK requires estimating the density function at zero, and ZKL requires  the density estimation at $K^\ast$
  quantiles. When the dimensionality is high, the estimation errors cause the empirical coverage rate of their methods to fall significantly below the nominal confidence level. In fact, the choice of the hyperparameter $h$ in the methods (YWZ,   ZKL and BCK) is critical.} Lastly, we observe that CRR-SCAD in general slightly outperforms CRR-LASSO, since its coverage is typically more accurate and the corresponding average lengths are narrower compared to using CRR-LASSO.

\section{Real data analysis}\label{sec:realdata}

In this section, we apply our proposed method to Cardiomyopathy microarray data (\cite{redfern1999conditional})  from a transgenic mouse model of dilated cardiomyopathy, which was analyzed by \cite{li2012feature}, \cite{feng2013} and \cite{cai2023tests}. This dataset is available on \url{https://sites.psu.edu/ril4/statistical-foundations-of-data-science/}, {\color{black} specifically under ``The Cardiomyopathy Microarray Data Used in Exercise 8.8" in F.Data sets for Chapter 8.}  It consists of $n
= 30$ observations
of mice, and the response variable  is  the  $\mbox{G}$ protein-coupled receptor $\mbox{Ro1}$ expression level. Besides, $p = 6319$ covariates measure the expression level of $6319$ genes. Since the dimension $p$ is very large compared with the sample size $n$,  the goal is to establish a high-dimensional linear model  and identify the influential genes for overexpression of the $\mbox{Ro1}$ in mice.   Following the previous literatures, both the response and predictors are standardized to be zero mean and unit variance before modeling.

Since it is expected that only a few of these genes are
related to the  $\mbox{Ro1}$ expression level, we first  screen the covariates using   the Kendall's $\tau$ correlation based screening  procedure developed by \cite{li2012robust} to select genes that possibly have effects on  the response, and  $50$ genes are picked out. To proceed, we apply the proposed debiased convoluted rank regression and the simultaneous inference procedure with $\mathcal G=\{1,\ldots,p\}$, and  identify two important genes  $\mbox{G3758}$ and $\mbox{G2375}$ at the significance  level of $\alpha=0.05$. The $95\%$ simultaneous
  confidence intervals  of the corresponding coefficients are $(0.180,1.157)$ for $\mbox{G3758}$, and $(0.133,1.110)$ for $\mbox{G2375}$. {\color{black}
  According to \cite{segal2003regression}, $\mbox{G3758}$   regulates the expression of the CD98 heavy chain,  a type II transmembrane glycoprotein that plays critical roles in amino acid transport, cell adhesion, and signaling; while $\mbox{G2375}$    controls the expression of Ribophorin II,   a transmembrane protein that plays a key role in  protein stability, folding, and proper function. It is worth noting that these two genes are also selected by the least angle regression and the support vector machines in \cite{segal2003regression}. Besides, 
   the genes selected by our proposed procedure align with those identified by \cite{li2012feature} using distance correlation screening, which again indicates the validity of our method. 
 Additionally, four relevant genes: $\mbox{G3758}$, $\mbox{G5940}$, $\mbox{G6031}$, and $\mbox{G4426}$, were identified using the inference method developed in \cite{yan2023confidence} with $\tau = 0.5$, which targets the median of the conditional distribution, further confirming the importance of $\mbox{G3758}$.
  However, the simultaneous inference procedures proposed in \cite{zhang2017simultaneous} and \cite{zhao2014general} fail to identify any significant predictor. }

Additionally, in order to further illustrate the robustness of the developed methods, we add some random outliers to the response variable. To be more specific, we randomly choose $10\%$ of the sample and add  random noise, which follows the Cauchy distribution with density $\left\{\pi c\left(1+x^2/c^2\right)\right\}^{-1}$, to the original response. Corresponding results 
  are reported in Table \ref{table5}. From the table, we can see that the results of the noised data are similar to those of the original data.  With the consistent findings, we have more  confidence in the relevance of the genes identified through our proposed procedures, providing valuable insights for future biological investigation.

\section{Conclusions and discussions}\label{sec6}
In this paper, we develop inference procedures for high-dimensional convoluted rank regression. We first obtain estimation error bounds of the penalized convoluted rank regression estimators under weaker conditions on the predictors.
We also introduce a debiased estimator. We establish Bahadur representation for our proposed estimator. We further develop simultaneous inference procedures. A novel bootstrap procedure is proposed and its validity is also established.

There are several possible topics for future research. Firstly, we may consider to investigate high-dimensional semiparametric convoluted rank regression. Secondly, we may consider high-dimensional convoluted rank regression in distributed data setting and online data setting. {\color{black}Lastly, in the high dimensional setting, \cite{fan2020comment} discussed the inference problem based on the rank lasso of \cite{wang2020tuning}. By an heuristic argument, we could conclude that 
the debiased convoluated rank regression estimator developed in this paper has the same asymptotic variance as the debiased rank regression estimator \citep{fan2020comment} when $h\rightarrow 0$. However, we should note that a formal proof needs more efforts due to the non-smoothness of the rank loss and is out scope of the current paper.
We will explore these possible topics in near future. }

\section*{Disclosure Statement}
The authors report there are no competing interests to declare.

{ \baselineskip 17pt
\bibliographystyle{apalike}
\bibliography{ref}
}

\newpage

\begin{table}[htbp]
 \centering
  \caption{Empirical coverage rates (CR) and average lengths (AL) of simultaneous confidence intervals for $\beta_k$, $k\in\mathcal G$  at the confidence
level $95\%$ when the error follows the standard normal distribution. }
\resizebox{\textwidth}{!}{
    \begin{tabular}{cccccccccccccc}
    \toprule
          & $\mathbf\Sigma$ & \multicolumn{6}{c}{Toeplitz}                  & \multicolumn{6}{c}{Block Diagonal} \\
\cmidrule{2-14}          & $\mathcal G$     & \multicolumn{2}{c}{$\{1,2,3,4,5\}$} & \multicolumn{2}{c}{$\{1,\ldots,[p/5]\}$} & \multicolumn{2}{c}{$\{1,\ldots,p\}$} & \multicolumn{2}{c}{$\{1,2,3,4,5\}$} & \multicolumn{2}{c}{$\{1,\ldots,[p/5]\}$} & \multicolumn{2}{c}{$\{1,\ldots,p\}$} \\
\cmidrule{2-14}    Method & $p$     & CR    & AL    & CR    & AL    & CR    & AL    & CR    & AL    & CR    & AL    & CR    & AL \\
    \midrule
    \multirow{3}[1]{*}{CRR-LASSO} & 50    & 0.958  & 0.786  & 0.960  & 0.859  & 0.974  & 1.016  & 0.938  & 0.893  & 0.954  & 0.996  & 0.972  & 1.194  \\
          & 100   & 0.940  & 0.833  & 0.970  & 0.984  & 0.976  & 1.140  & 0.924  & 0.946  & 0.952  & 1.140  & 0.978  & 1.327  \\
          & 200   & 0.936  & 0.887  & 0.964  & 1.118  & 0.976  & 1.279  & 0.896  & 0.969  & 0.950  & 1.248  & 0.974  & 1.434  \\
    \multirow{3}[0]{*}{CRR-SCAD} & 50    & 0.970  & 0.719  & 0.950  & 0.799  & 0.974  & 0.955  & 0.976  & 0.812  & 0.970  & 0.920  & 0.970  & 1.115  \\
          & 100   & 0.970  & 0.746  & 0.968  & 0.904  & 0.974  & 1.055  & 0.958  & 0.829  & 0.978  & 1.028  & 0.968  & 1.209  \\
          & 200   & 0.968  & 0.772  & 0.972  & 1.016  & 0.974  & 1.165  & 0.974  & 0.833  & 0.976  & 1.117  & 0.978  & 1.291  \\
    \multirow{3}[0]{*}{ZC} & 50    & 0.928  & 0.635  & 0.948  & 0.700  & 0.980  & 0.826  & 0.874  & 0.670  & 0.908  & 0.742  & 0.950  & 0.879  \\
          & 100   & 0.880  & 0.628  & 0.926  & 0.749  & 0.952  & 0.866  & 0.866  & 0.651  & 0.912  & 0.781  & 0.956  & 0.906  \\
          & 200   & 0.838  & 0.603  & 0.908  & 0.769  & 0.916  & 0.873  & 0.808  & 0.614  & 0.900  & 0.787  & 0.928  & 0.895  \\
    \multirow{3}[0]{*}{YWZ} & 50    & 0.876  & 1.056  & 0.902  & 1.154  & 0.884  & 1.356  & 0.824  & 1.064  & 0.860  & 1.161  & 0.862  & 1.368  \\
          & 100   & 0.938  & 1.593  & 0.972  & 1.867  & 0.968  & 2.151  & 0.908  & 1.612  & 0.960  & 1.893  & 0.964  & 2.181  \\
          & 200   & 0.902  & 2.079  & 0.980  & 2.606  & 0.990  & 2.954  & 0.916  & 2.138  & 0.984  & 2.676  & 0.998  & 3.041  \\
    \multirow{3}[0]{*}{ZKL} & 50    & 0.844  & 0.595  & 0.866  & 0.658  & 0.844  & 0.788  & 0.866  & 0.692  & 0.894  & 0.777  & 0.940  & 0.945  \\
          & 100   & 0.774  & 0.519  & 0.784  & 0.628  & 0.752  & 0.741  & 0.752  & 0.585  & 0.832  & 0.720  & 0.852  & 0.863  \\
          & 200   & 0.446  & 0.381  & 0.482  & 0.502  & 0.392  & 0.588  & 0.464  & 0.413  & 0.532  & 0.555  & 0.534  & 0.657  \\
    \multirow{3}[0]{*}{ZKL*} & 50    & 0.934  & 0.695  & 0.938  & 0.768  & 0.936  & 0.921  & 0.944  & 0.803  & 0.954  & 0.902  & 0.980  & 1.097  \\
          & 100   & 0.922  & 0.704  & 0.964  & 0.851  & 0.964  & 1.004  & 0.934  & 0.800  & 0.972  & 0.985  & 0.984  & 1.179  \\
          & 200   & 0.908  & 0.720  & 0.948  & 0.948  & 0.986  & 1.111  & 0.912  & 0.785  & 0.962  & 1.056  & 0.978  & 1.248  \\
    \multirow{3}[1]{*}{BCK} & 50    & 0.812  & 0.701  & 0.866  & 0.773  & 0.952  & 0.915  & 0.822  & 0.933  & 0.880  & 1.110  & 0.948  & 1.382  \\
          & 100   & 0.798  & 0.624  & 0.882  & 0.747  & 0.936  & 0.860  & 0.806  & 0.804  & 0.888  & 1.058  & 0.932  & 1.244  \\
          & 200   & 0.558  & 0.518  & 0.764  & 0.670  & 0.796  & 0.763  & 0.548  & 0.660  & 0.740  & 0.903  & 0.780  & 1.035  \\
    \bottomrule
    \end{tabular}%
    }
  \label{table1}%
\end{table}%

\begin{table}[htbp]
 \centering
  \caption{Empirical coverage rates (CR) and average lengths (AL) of simultaneous confidence intervals for $\beta_k$, $k\in\mathcal G$  at the confidence
level $95\%$ when the error follows the mixture  normal distribution $0.95N(0,1)+0.05N(0,100^2)$.}
    \resizebox{\textwidth}{!}{
    \begin{tabular}{cccccccccccccc}
    \toprule
          & $\mathbf\Sigma$ & \multicolumn{6}{c}{Toeplitz}                  & \multicolumn{6}{c}{Block Diagonal} \\
\cmidrule{2-14}          & $\mathcal G$     & \multicolumn{2}{c}{$\{1,2,3,4,5\}$} & \multicolumn{2}{c}{$\{1,\ldots,[p/5]\}$} & \multicolumn{2}{c}{$\{1,\ldots,p\}$} & \multicolumn{2}{c}{$\{1,2,3,4,5\}$} & \multicolumn{2}{c}{$\{1,\ldots,[p/5]\}$} & \multicolumn{2}{c}{$\{1,\ldots,p\}$} \\
\cmidrule{2-14}    Method & p     & CR    & AL    & CR    & AL    & CR    & AL    & CR    & AL    & CR    & AL    & CR    & AL \\
    \midrule
    \multirow{3}[1]{*}{CRR-LASSO} & 50    & 0.956  & 0.898  & 0.960  & 0.982  & 0.974  & 1.160  & 0.948  & 1.039  & 0.948  & 1.160  & 0.970  & 1.390  \\
          & 100   & 0.954  & 1.005  & 0.970  & 1.181  & 0.978  & 1.372  & 0.932  & 1.126  & 0.948  & 1.352  & 0.972  & 1.580  \\
          & 200   & 0.934  & 1.129  & 0.970  & 1.427  & 0.978  & 1.631  & 0.880  & 1.311  & 0.958  & 1.676  & 0.978  & 1.926  \\
    \multirow{3}[0]{*}{CRR-SCAD} & 50    & 0.968  & 0.806  & 0.974  & 0.896  & 0.972  & 1.071  & 0.968  & 0.917  & 0.962  & 1.039  & 0.974  & 1.257  \\
          & 100   & 0.966  & 0.844  & 0.972  & 1.022  & 0.976  & 1.196  & 0.962  & 0.945  & 0.976  & 1.166  & 0.976  & 1.375  \\
          & 200   & 0.964  & 0.876  & 0.972  & 1.145  & 0.976  & 1.316  & 0.966  & 0.963  & 0.968  & 1.273  & 0.968  & 1.469  \\
    \multirow{3}[0]{*}{ZC} & 50    & 0.946  & 13.560  & 0.948  & 14.946  & 0.968  & 17.605  & 0.936  & 14.224  & 0.952  & 15.715  & 0.980  & 18.634  \\
          & 100   & 0.922  & 12.818  & 0.960  & 15.283  & 0.966  & 17.670  & 0.934  & 13.329  & 0.946  & 15.973  & 0.964  & 18.489  \\
          & 200   & 0.926  & 12.876  & 0.948  & 16.440  & 0.964  & 18.672  & 0.938  & 13.063  & 0.956  & 16.744  & 0.958  & 19.066  \\
    \multirow{3}[0]{*}{YWZ} & 50    & 0.878  & 1.097  & 0.882  & 1.201  & 0.874  & 1.407  & 0.870  & 1.119  & 0.884  & 1.221  & 0.878  & 1.437  \\
          & 100   & 0.944  & 1.635  & 0.968  & 1.912  & 0.964  & 2.206  & 0.900  & 1.683  & 0.950  & 1.969  & 0.966  & 2.273  \\
          & 200   & 0.902  & 2.269  & 0.984  & 2.826  & 0.996  & 3.204  & 0.886  & 2.327  & 0.990  & 2.905  & 0.998  & 3.300  \\
    \multirow{3}[0]{*}{ZKL} & 50    & 0.882  & 0.688  & 0.876  & 0.765  & 0.860  & 0.916  & 0.906  & 0.802  & 0.916  & 0.904  & 0.944  & 1.093  \\
          & 100   & 0.772  & 0.616  & 0.840  & 0.753  & 0.806  & 0.883  & 0.806  & 0.702  & 0.868  & 0.870  & 0.902  & 1.041  \\
          & 200   & 0.566  & 0.461  & 0.614  & 0.609  & 0.562  & 0.714  & 0.538  & 0.500  & 0.628  & 0.673  & 0.616  & 0.797  \\
    \multirow{3}[0]{*}{ZKL*} & 50    & 0.922  & 0.751  & 0.910  & 0.835  & 0.932  & 1.000  & 0.940  & 0.873  & 0.960  & 0.984  & 0.974  & 1.190  \\
          & 100   & 0.904  & 0.762  & 0.940  & 0.932  & 0.946  & 1.093  & 0.930  & 0.864  & 0.962  & 1.071  & 0.992  & 1.283  \\
          & 200   & 0.902  & 0.773  & 0.956  & 1.022  & 0.966  & 1.198  & 0.898  & 0.844  & 0.962  & 1.136  & 0.986  & 1.344  \\
    \multirow{3}[1]{*}{BCK} & 50    & 0.840  & 0.765  & 0.866  & 0.848  & 0.928  & 1.002  & 0.868  & 1.030  & 0.906  & 1.228  & 0.962  & 1.525  \\
          & 100   & 0.768  & 0.676  & 0.854  & 0.805  & 0.882  & 0.933  & 0.758  & 0.887  & 0.860  & 1.149  & 0.926  & 1.349  \\
          & 200   & 0.626  & 0.576  & 0.794  & 0.746  & 0.850  & 0.844  & 0.660  & 0.711  & 0.798  & 1.006  & 0.862  & 1.150  \\
    \bottomrule
    \end{tabular}%
    }
  \label{table2}%
\end{table}%

\begin{table}[htbp]
   \centering
  \caption{Empirical coverage rates (CR) and average lengths (AL) of simultaneous confidence intervals for $\beta_k$, $k\in\mathcal G$  at the confidence
level $95\%$ when the error follows the standard  Cauchy distribution.}
    \resizebox{\textwidth}{!}{
    \begin{tabular}{cccccccccccccc}
    \toprule
          & $\mathbf\Sigma$ & \multicolumn{6}{c}{Toeplitz}                  & \multicolumn{6}{c}{Block Diagonal} \\
\cmidrule{2-14}          & $\mathcal G$     & \multicolumn{2}{c}{$\{1,2,3,4,5\}$} & \multicolumn{2}{c}{$\{1,\ldots,[p/5]\}$} & \multicolumn{2}{c}{$\{1,\ldots,p\}$} & \multicolumn{2}{c}{$\{1,2,3,4,5\}$} & \multicolumn{2}{c}{$\{1,\ldots,[p/5]\}$} & \multicolumn{2}{c}{$\{1,\ldots,p\}$} \\
\cmidrule{2-14}    Method & p     & CR    & AL    & CR    & AL    & CR    & AL    & CR    & AL    & CR    & AL    & CR    & AL \\
    \midrule
    \multirow{3}[1]{*}{CRR-LASSO} & 50    & 0.944  & 1.665  & 0.958  & 1.830  & 0.974  & 2.164  & 0.942  & 1.971  & 0.948  & 2.191  & 0.974  & 2.614  \\
          & 100   & 0.924  & 1.910  & 0.956  & 2.258  & 0.968  & 2.614  & 0.922  & 2.229  & 0.960  & 2.697  & 0.972  & 3.148  \\
          & 200   & 0.916  & 2.546  & 0.962  & 3.236  & 0.970  & 3.709  & 0.898  & 2.890  & 0.940  & 3.735  & 0.970  & 4.308  \\
    \multirow{3}[0]{*}{CRR-SCAD} & 50    & 0.944  & 1.376  & 0.956  & 1.530  & 0.974  & 1.827  & 0.948  & 1.644  & 0.954  & 1.845  & 0.970  & 2.220  \\
          & 100   & 0.942  & 1.513  & 0.960  & 1.828  & 0.964  & 2.126  & 0.942  & 1.746  & 0.954  & 2.147  & 0.968  & 2.516  \\
          & 200   & 0.940  & 1.676  & 0.960  & 2.178  & 0.964  & 2.502  & 0.938  & 1.910  & 0.948  & 2.503  & 0.968  & 2.892  \\
    \multirow{3}[0]{*}{ZC} & 50    & 0.920  & 68.510  & 0.940  & 75.795  & 0.966  & 87.666  & 0.936  & 70.994  & 0.952  & 80.876  & 0.966  & 94.488  \\
          & 100   & 0.920  & 60.488  & 0.946  & 76.075  & 0.966  & 88.004  & 0.932  & 68.781  & 0.958  & 80.071  & 0.964  & 91.899  \\
          & 200   & 0.926  & 300.217  & 0.964  & 381.406  & 0.962  & 438.606  & 0.904  & 306.019  & 0.940  & 399.025  & 0.968  & 447.207  \\
    \multirow{3}[0]{*}{YWZ} & 50    & 0.844  & 1.591  & 0.850  & 1.716  & 0.886  & 2.003  & 0.846  & 1.668  & 0.874  & 1.812  & 0.906  & 2.125  \\
          & 100   & 0.904  & 2.176  & 0.968  & 2.532  & 0.968  & 2.921  & 0.859  & 2.271  & 0.950  & 2.657  & 0.968  & 3.056  \\
          & 200   & 0.860  & 2.997  & 0.970  & 3.704  & 0.994  & 4.184  & 0.816  & 3.014  & 0.982  & 3.738  & 0.988  & 4.245  \\
    \multirow{3}[0]{*}{ZKL} & 50    & 0.878  & 1.096  & 0.856  & 1.216  & 0.864  & 1.458  & 0.898  & 1.266  & 0.912  & 1.429  & 0.944  & 1.745  \\
          & 100   & 0.768  & 0.996  & 0.796  & 1.207  & 0.774  & 1.423  & 0.798  & 1.135  & 0.836  & 1.413  & 0.874  & 1.689  \\
          & 200   & 0.542  & 0.765  & 0.612  & 1.009  & 0.540  & 1.180  & 0.590  & 0.822  & 0.642  & 1.117  & 0.650  & 1.317  \\
    \multirow{3}[0]{*}{ZKL*} & 50    & 0.896  & 1.165  & 0.882  & 1.292  & 0.924  & 1.549  & 0.930  & 1.341  & 0.954  & 1.513  & 0.972  & 1.847  \\
          & 100   & 0.876  & 1.164  & 0.902  & 1.411  & 0.910  & 1.664  & 0.880  & 1.318  & 0.932  & 1.640  & 0.952  & 1.960  \\
          & 200   & 0.854  & 1.146  & 0.900  & 1.517  & 0.910  & 1.771  & 0.842  & 1.245  & 0.932  & 1.691  & 0.946  & 1.992  \\
    \multirow{3}[1]{*}{BCK} & 50    & 0.914  & 1.362  & 0.952  & 1.503  & 0.970  & 1.781  & 0.916  & 1.828  & 0.950  & 2.197  & 0.978  & 2.709  \\
          & 100   & 0.854  & 1.253  & 0.916  & 1.491  & 0.950  & 1.726  & 0.898  & 1.683  & 0.944  & 2.207  & 0.972  & 2.606  \\
          & 200   & 0.738  & 1.081  & 0.890  & 1.392  & 0.930  & 1.587  & 0.758  & 1.359  & 0.872  & 1.898  & 0.926  & 2.173  \\
    \bottomrule
    \end{tabular}%
    }
  \label{table3}%
\end{table}%

\begin{table}[htbp]
  \centering
  \caption{The scale parameter $c$ of the artificial noise, IDs of the  identified genes, and  $95\%$ simultaneous confidence intervals (SCIs) along with $95\%$   confidence intervals (CIs) of the significant  coefficients.}
    \begin{tabular}{ccccccc}
    \toprule
    c     & ID    & $95\%$  SCI & $95\%$ CI & ID    & $95\%$ SCI & $95\%$ CI \\
    \midrule
    0     & G3758 & (0.180,1.157) & (0.275,1.062) & G2375 & (0.133,1.110) & (0.266,0.977) \\
    20    & G3758 & (0.226,1.130) & (0.314,1.041) & G2375 & (0.134,1.037) & (0.255,0.916) \\
    40    & G3758 & (0.244,1.150) & (0.332,1.062) & G2375 & (0.122,1.029) & (0.247,0.904) \\
    60    & G3758 & (0.256,1.188) & (0.338,1.106) & G2375 & (0.113,1.044) & (0.237,0.921) \\
    80    & G3758 & (0.207,1.091) & (0.284,1.013) & G2375 & (0.139,1.023) & (0.256,0.906) \\
    100   & G3758 & (0.209,1.092) & (0.283,1.018) & G2375 & (0.137,1.019) & (0.250,0.906) \\
    \bottomrule
    \end{tabular}%
  \label{table5}%
\end{table}%

\end{document}